\documentclass{eas}
\usepackage{graphicx}

\begin{document}

%%-----------------------------
%%      the top matter
%%-----------------------------
\title{Modelling binary rotating stars by new population synthesis code BONNFIRES} 
\runningtitle{BONNFIRES}
\author{Herbert H.B. Lau}\address{Argelander-Instit\"ut fur Astronomie, Universit\"at Bonn, Auf dem H\"ugel 71, 53121 Bonn; \email{hblau@astro.uni-bonn.de}}
\author{Robert G. Izzard}\sameaddress{1}
\author{Fabian R.N. Schneider}\sameaddress{1}

\begin{abstract}
BONNFIRES, a new generation of population synthesis code, can calculate nuclear reaction, various mixing processes
and binary interaction in a timely fashion.  We use this new population synthesis code to
study the interplay between binary mass transfer and rotation. We 
aim to compare theoretical models with observations, in particular the surface nitrogen
abundance and rotational velocity. Preliminary results show binary
interactions may explain the formation of nitrogen-rich slow rotators
and nitrogen-poor fast rotators, but more work needs to be done to estimate whether the observed frequencies of those stars can be matched.
\end{abstract}
\maketitle
%%-----------------------------
%%      your text
%%-----------------------------
\section{Introduction}
Rapid rotation can lead to efficient mixing of the whole star and hence alter the nucleosynthesis process and observed abundances. Hydrogen burning products, such as nitrogen, are a strong indicator of rotational mixing, so nitrogen are expected to be proportional to observed rotational velocities. How the evolution and final fate of the star depends on rotation rate has been recently reviewed by Langer (\cite{Langer2012}) and Maeder \& Meynet (\cite{Maeder2012}).

However, observations show that the picture of rotational mixing in massive stars are not that simple. Hunter {\em et al.\/} (\cite{Hunter2008}) found a significant fraction of stars that cannot be explained by current single rotating stars models. These are stars with strong nitrogen enhancement without rapid rotations and stars with rapid rotation without nitrogen enhancement. These observations show that other processes must be responsible for either the nitrogen enhancement or the current rotation velocity. 

Sana {\em et al.\/} (\cite{Sana2013}) shows that almost all massive stars are born in a binary system , so significant fractions of stars observed must have undergone binary interaction to reach their current states. The aim of this work is to investigate whether binary interactions can account for the observed anomalies between nitrogen abundances and rotational velocities.

\begin{figure*}[h]
\resizebox{\hsize}{!}{\includegraphics[angle=270,clip=true]{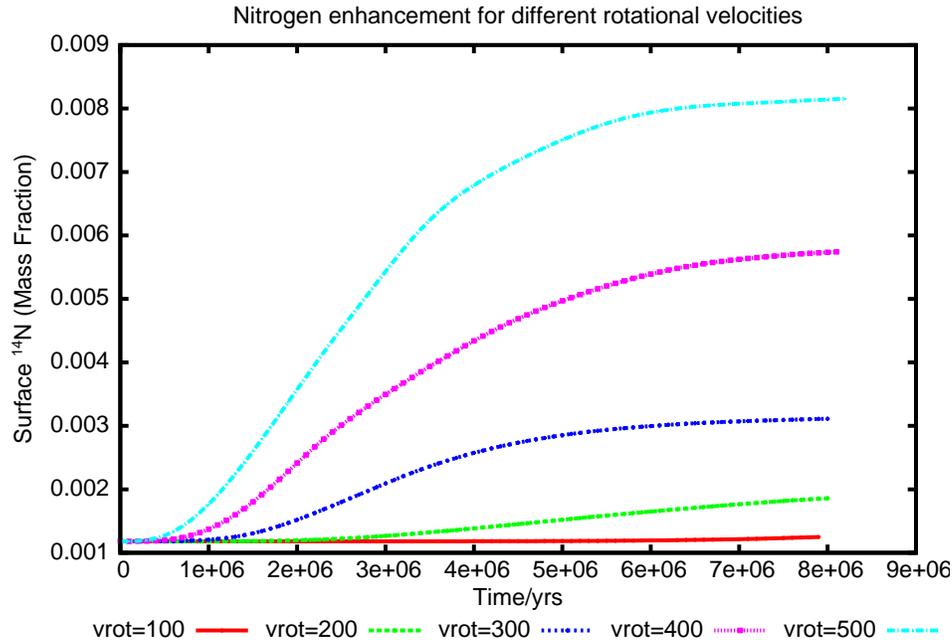}}
\caption{\footnotesize
Evolution of surface nitrogen abundances (by mass fraction) of $20 M_{\odot}$ models at solar metallicity (Z=0.02) with different surface rotational velocity from $100-500 \rm{km s}^{-1}$. Models are produced by the population synthesis code BONNFIRES and stop at the end of main sequence. Fast rotators enhance significantly more nitrogen with longer main sequence lifetimes.
}
\label{fig:ZM}
\end{figure*}

\section{New Population synthesis code BONNFIRES}
BONNFIRES is designed to model the internal mixing and nucleosynthesis of binary stars in a timely fashion. Internal structure variables are interpolated from input models produced by detailed evolutionary code. Binary physics, mixing and nuclear reaction are calculated independently by BONNFIRES to study binary interactions and internal mixing. This will be the first population synthesis code with detailed internal composition profile, mixing processes and nuclear reaction networks. We follow Heger {\em et al.\/} (\cite{Heger}) for treatments of rotational mixing and Brott {\em et al.\/} (\cite{Brott2011}) for rotational mixing parameters. Evolution of the binary systems are followed till the end of the main-sequence of the primary stars.

Figure 1 shows $20 M_{\odot}$ models with different initial rotational velocities by BONNFIRES. The surface nitrogen enhancement is much stronger for fast rotators. Moreover, the main sequence lifetimes for fast rotators are longer because of additional hydrogen that are mixed down to the burning region. This result is consistent with models made by detailed evolutionary codes.

\begin{figure*}[h]
\resizebox{\hsize}{!}{\includegraphics[angle=0,clip=true]{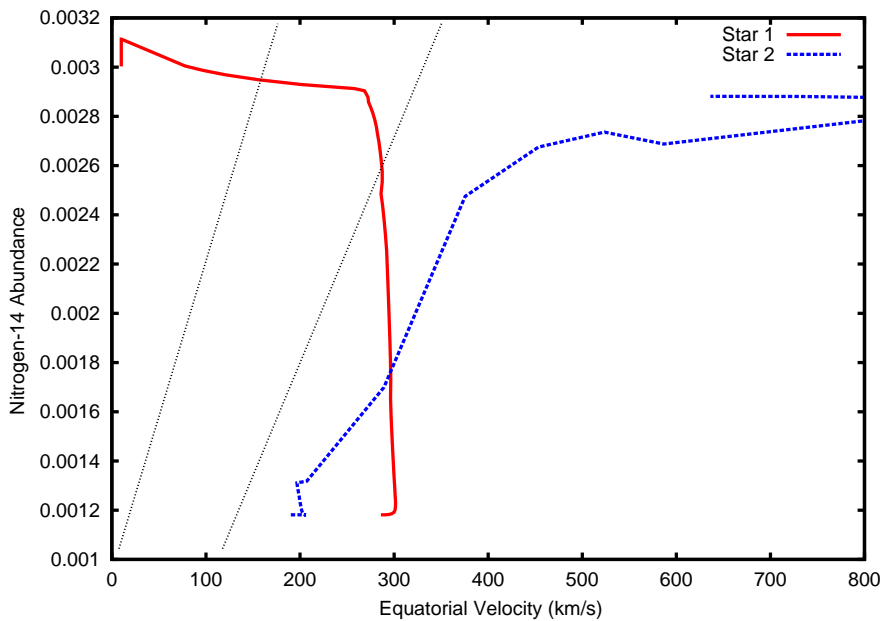}}
\caption{\footnotesize
Evolution of surface nitrogen abundances (by mass fraction) against equatorial rotational velocities for a binary system consisting of a $16 M_{\odot}$ primary, a $12 M_{\odot}$ secondary with 3 days period. The initial rotational velocity for primary and secondary are $300\rm{km s}^{-1}$ and $200\rm{km s}^{-1}$ respectively. Region between two dotted black lines are what is expected from single star models. After mass transfer, the nitrogen-rich primary (red solid line) spins down  while the secondary (blue dashed line) spins up but with significantly less nitrogen enrichment compared to single star models.
}
\label{fig:Hunter}
\end{figure*}

\section{Result}

For the short period systems, the primary stars interact with their companions during the main-sequence. Typically, when the primary overflow its Roche Lobe, the star is expanding and hence spin down after mass loss. If the primary is initially rapidly rotating, its surface will be rich in nitrogen but it is now a slow rotating after mass transfer. Alternatively, if the primary suffers heavy mass loss, its inner hydrgeon burning region will be exposed and so the surface will now be rich in nitrogen irregardless of any prior rotational mixing. These are possible formation channels of nitrogen-rich slow rotators.

Materials transferred onto companion stars also carry angular momentum, so the companion stars will be spun up and rotate faster. The surface abundances of the secondary depends on the compositions of the accreted materials as well as any subsequent mixing. Because the secondary is now rapidly rotating it may mix materials from its inner hydrogen-burning region to the surface. However, if the secondary is previously slow rotating, the burning region has a much higher mean molecular weight. This mean molecular weight barrier can suppress rotational mixing, so nitrogen enhancement is much smaller than its single star counterpart that rotates at the same velocity from the beginning. It is therefore possible the secondary is observed as nitrogen-poor fast rotators.

Figure 2 illustrates a typical short period binary system. This systems consists of a $16 M_{\odot}$ primary, a $12 M_{\odot}$ secondary with 3 days period. The primary is initially rotating at $300\rm{km\,s}^{-1}$ and enhanced surface nitrogen to $XN=0.029$ through rotational mixing, similar to a single star. Then mass transfer from primary to secondary occurs and the primary spins down rapidly to become a nitrogen-rich slow rotators. Surface nitrogen abundance of secondary also increase because the accreted material is enriched in nitrogen, but the secondary is spun up to critical velocity. If we compared to single stars model with the same rotational velocities, the secondary has significantly less nitrogen enhancement on the surface. 

\section{Future Work}
In order to statistically compare with observations, we need to take into account the frequency of the above channels and also for how long the stars can be observed nitrogen-rich slow rotators and nitrogen-poor fast rotators. We will produce a fine grid of binary models with various assumptions of tides, efficiency of rotational mixing to determine whether the observed frequency can be reproduced. Results should be expected in the early 2014.

%%-----------------------------
%%      your bibliography
%%-----------------------------
%%\bibliography{HBL}

\end{document}